\documentclass[onecolumn,prb,english]{revtex4-1}
\usepackage[latin9]{inputenc}
\usepackage{amstext,bm}
\usepackage{graphicx}
\usepackage{xcolor}
\usepackage{color}
\usepackage{amsmath}
\usepackage{babel}
\usepackage[titletoc]{appendix}
\usepackage{hyperref}
\hypersetup{
     colorlinks = true,
     citecolor  = magenta,  
     linkcolor  = magenta 
}
\usepackage{float}

\begin{document} 
\title{Topological nodal states in circuit lattice} 

\author{Kaifa Luo$^{1}$}
\author{Rui Yu$^{1}$}
\email{yurui1983@foxmail.com}
\author{Hongming Weng$^{2}$}
\email{hmweng@iphy.ac.cn}
\affiliation{$^{1}$ 
School of Physics and Technology, Wuhan University, Wuhan 430072,China}
\affiliation{$^{2}$
Beijing National Laboratory for Condensed Matter Physics, and Institute of
Physics, Chinese Academy of Sciences, Beijing 100190, China}

\begin{abstract}
The search for artificial structure with tunable topological properties is an interesting research direction of today's topological physics. 
Here, we introduce a scheme to realize `topological semimetal states' with
a three-dimensional periodic inductor-capacitor (LC) circuit lattice,
where the topological nodal-line state and Weyl state can be achieved by 
tuning the parameters of inductors and capacitors.
A tight-binding-like model is derived to analyze the topological properties of
the LC circuit lattice. 
The key characters of the topological states, such as the drumhead-like 
surface bands for nodal-line state
and the Fermi-arc-like surface bands for Weyl state, are found in these systems. 
We also show that the Weyl points are stable with the fabrication errors of electric devices.
\end{abstract}
\maketitle

Recently, there are great interesting in realizing topological 
states in various platforms. 
Topological states, including the quantum
Hall states, quantum spin Hall states, Dirac states, Weyl states and nodal-line states etc., have achieved significant progresses in 
electronic materials\cite{TI_RMP_2010,TI_RMP_Qi2011,Weyl_WXG_PRL_2011,Weyl_XuGang_PRL_2011,Weyl_HMWeng_PRX_2015,Weyl_NC_2015,QAHE_Rev_AP_2015,Weyl_EXP_PRX_2015,Weyl_EXP_S_2015,Weyl_review_2017},
cold atoms\cite{ColdAtom_goldman_realistic_2010,ColdAtom_sun_topological_2012,ColdAtom_jotzu_HaldaneModel_2014,Weyl_ColdAtom_ZhangChuanwei_PRL_2015,ColdAtom_aidelsburger_ChernNumber_Hofstadter_2015,Weyl_ColdAtom_PRL_2015}, 
photonics\cite{Photonic_Haldane_analogs_2008,Photonic_haldane_OpticalWaveguides_2008,PhotonicTI_LingLu_NPgyroid_2013,Photonic_rechtsman_photonic_2013,PhotonicTI_LingLu_NP_2014}, 
phononics\cite{
Topo_phonon_prodan_topological_2009,
Topo_phonon_berg_topological_2011,
Topo_phonon_wang_topological_2015,
Topo_phonon_po_phonon_2016,
Topo_phonon_xiao_topological_2017,
Weyl_phono_ZYLiu_NP_2017,
Topo_phonon_zhang_double_weyl_2017}, 
and mechanical systems\cite{
topo_mecha_kane_topological_2013,
topo_mecha_chen_nonlinear_2014,
topo_mecha_susstrunk_observation_2015,
topo_mecha_wang_coriolis_2015,
topo_mecha_nash_topological_2015,
topo_mecha_paulose_topological_2015,
topo_mecha_kariyado_manipulation_2015,
topo_mecha_rocklin_directional_2016,
topo_mecha_xiong_effects_2016,
topo_mecha_susstrunk_classification_2016,
topo_mecha_meeussen_geared_2016,
topo_mecha_rocklin_mechanical_2016,
topo_mecha_chen_topological_2016,
topo_mecha_wang_elastic_2016,
topo_mecha_zhou_kink_antikink_2017,
topo_mecha_rocklin_transformable_2017,
topo_mecha_coulais_static_2017,
topo_mecha_takahashi_edge_2017}.
In addition, the topological properties in electric circuit system have also been explored in several works\cite{Circuit_2D_PRX_2015,Circuit_2D_PRL_2015,Circuit_floquet_TI_IEEE_2016,Circuit_3D_Arxiv_2017,Circuit_1D_Arxiv_2017,li_generating_2018,tongji_prb2018,hadad_self-induced_2018}. 
The quantum spin Hall-like states have been proposed in two-dimensional circuit lattice
via time-reversal symmetric Hofstadter model\cite{Circuit_2D_PRX_2015,Circuit_2D_PRL_2015}. 
The Weyl state has been found in three-dimensional circuit network and proposed to be able to be detected from the boundary resonant signal\cite{Circuit_3D_Arxiv_2017}.
Based on the linear and nonlinear one-dimensional SSH-type circuit arrays, the topological Zak phase\cite{Circuit_1D_Arxiv_2017} and self-induced topological properties\cite{hadad_self-induced_2018} have been discussed recently.
Most of these proposed electric circuits are composed of interconnected linear lossless 
passive elements, such as capacitors and inductors. 
One of the most significant advantage of the circuit lattice is that 
each parameter in the system is independently artificial adjustable 
and the symmetry of the lattice is protected by the parameters of
electronic components and the way they are connected, rather than their positions
in the real space.  

In the present work, we demonstrate a feasible strategy to design 
both nodal-line state and Weyl state in a three-dimensional
circuit lattice. 
The topological phase transition between them can be controlled by tuning the parameters of the components.
In order to investigate their topological properties,
we transform the circuit network problem to tight-binding-like model, 
based on which the novel surface states, including drumhead-like surface bands for nodal-line state and Fermi arc-like surface bands for Weyl state, are found in the surface of the circuit lattices. 
Moreover, the stability of the Weyl points under perturbation of fabrication errors are studied.

\section{Models and Theoretical Framework}

The design scheme of realizing the nodal-line state and Weyl state in LC circuit lattice are shown below. We consider a honeycomb lattice consisting of capacitors and inductors in $\bm{a}$-$\bm{b}$ plane as a starting point.
The sub-nodes A and B are linked by capacitors $C_{1}$, $C_{2}$ and $C_{3}$.  nodes $A$ ($B$) are grounded through the parallel connected inductor $L_{A}$ ($L_{B}$) and capacitor $C_{GA}$ ($C_{GB}$) as shown in Fig.~1(a).
Similar to the electronic band structure of graphene, the frequency spectrum of the single layer LC honeycomb lattice contains two band-crossing points in the two-dimensional Brillouin zone (BZ). 
Stacking the two-dimensional honeycomb lattice along $\bm c$ direction without any coupling between each other, the 
band-crossing points will form two straight nodal-lines in the three-dimensional BZ as shown in Fig.~1(a).
In order to make the straight nodal-lines are $\bm k_c$ dependent,
we connect nodes A and B between neighbor-layers with $C_4$ as shown in Fig.~1(b). 
By tuning the value of $C_4$, these two separate lines are deformed and merged into a closed ring, namely, the nodal-line we are searching for.
The nodal-line structure is protected by the coexistence of time
reversal and space inversion symmetry. 
If the space inversion symmetry of the circuit system is removed, the continuous nodal-line may be degenerated to discrete Weyl points\cite{PhotonicTI_LingLu_NPgyroid_2013}.
Following this insight, node pairs A-A and B-B between the nearest-neighbor-layers are connected with $C_A$ and $C_B$, respectively, as shown in Fig.~1(c). $C_A$ and $C_B$, $C_{GA}$ and $C_{GB}$ are deliberately set to be different, resulting in broken of space inversion symmetry and emergence of the Weyl points. 
The LC circuit lattice in Fig.~1(c) can be deformed to Fig.~1(d), which bring convenience to construction of circuit elements in experiments with spectrum topologically invariable.
The band structures and the topological properties of the nodal-line state and Weyl state for lattice given in Fig.~1(b) and Fig.~1(c, d) will be detailed in the remain text.

\begin{figure}[h]
\includegraphics[width=0.99\textwidth]{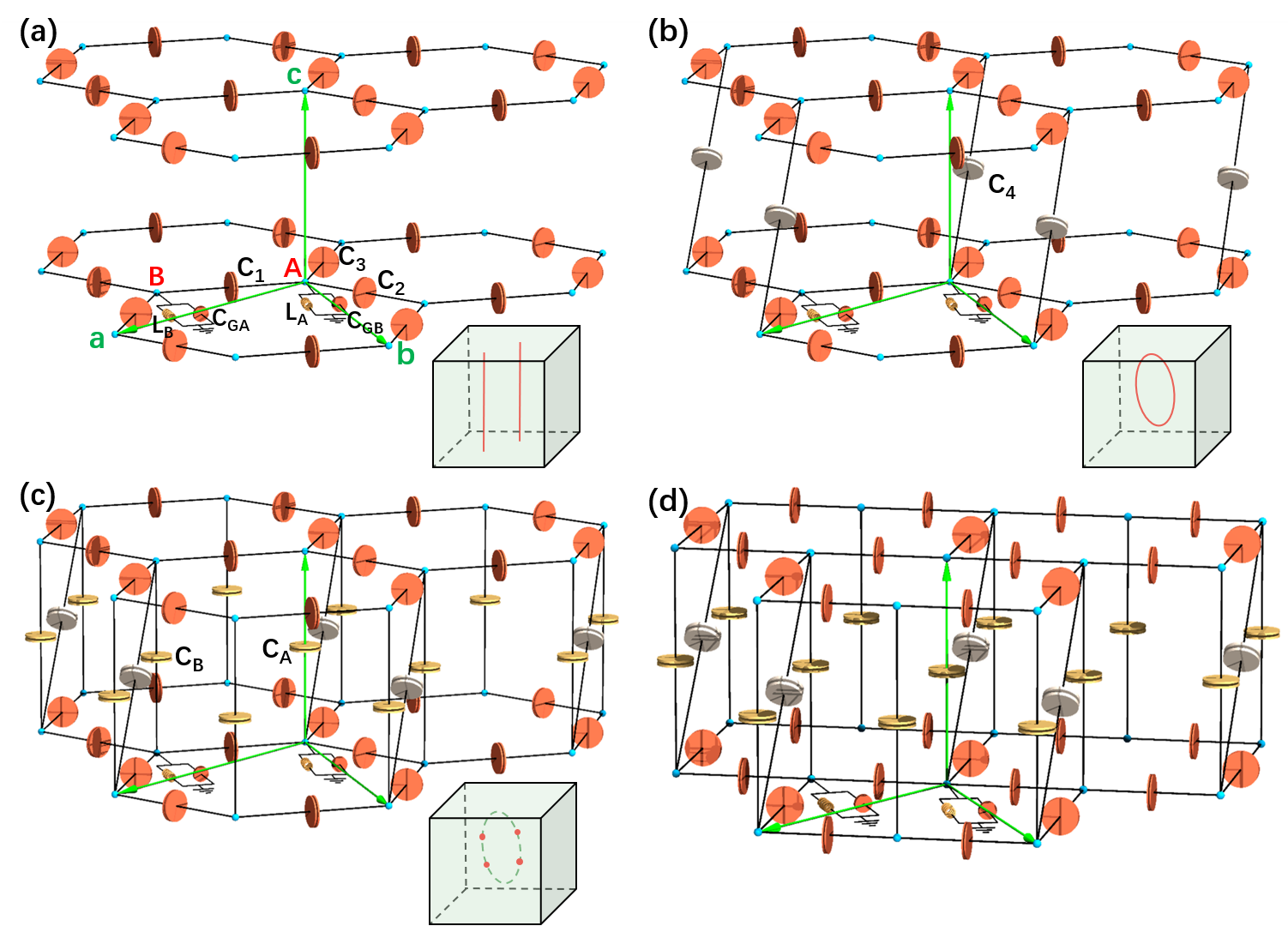}
\caption{Schematic setup of the three-dimensional LC circuit lattice. 
(a) Honeycomb layers consisting of inductors and capacitors stack along $\bm c$
direction without connection between each layer. 
The primitive unit cell consists of two inequivalent
nodes A and B, which are linked by capacitors $C_{1}$, $C_{2}$ and
$C_{3}$ in the $\bm a$-$\bm b$ plane. Each node $A$ ($B$) is grounded through
the parallel connected inductor $L_{A}$ ($L_{B}$) and capacitor
$C_{GA}$ ($C_{GB}$). Lattice vectors are denoted as $\bm{a},$
$\bm{b}$ and $\bm{c}$.
The frequency bands structure of the single layer honeycomb LC lattice has two band crossing points, which are uniform in $\bm{k_c}$ direction and form two straight nodal-lines in the BZ
(red color lines in the inset).
(b) Connecting nodes A and nodes B between neighbor-layers with $C_4$. The nodal-lines 
become $\bm k_c$ dependent and form a closed ring by choosing appropriate $C_4$.
(c) Connecting nodes A-A and nodes B-B between neighbor-layers with $C_A$ and $C_B$, respectively, and removing the space inversion symmetry by tuning $C_A \ne C_B$ and 
$C_{GA} \ne C_{GB}$, the nodal-ring may be degenerated to Weyl points. 
The LC lattice can be deformed into (d), 
which bring convenience for constructing circuit elements in experiments with spectrum topologically invariable.}
\end{figure}

In the present work, we study the resonance condition
of the circuit lattice, where a non-zero distribution
of potential satisfies Kirchhoff's law. 
We follow the method given in Ref.\cite{Crcuit_TB_PSS_2012}, where the periodical circuit lattice problem is transformed into a tight-binding-like model in the momentum 
space. This approach relies on an analogy 
between Kirchhoff current equation in periodic circuit lattice and the
quantum mechanics with periodic crystalline structure, from which the
circuit band structure arises in a manner analogous to electronic band 
structure in crystals.
The tight-binding-like model for the circuit lattice in Fig.~1(c,d)
is given as
\begin{equation}
H(\bm{k})|\mathcal V(\bm{k})\rangle=\frac{1}{\omega^{2}L}|\mathcal V(\bm{k})\rangle,\label{eq:Hpsi}
\end{equation}
where 
\begin{equation}
H(\bm{k})=\sum_{i=0}^{3}d_{i}(\bm{k})\sigma_{i},\label{eq:H_bulk}
\end{equation}
$\omega$ is the resonance frequency of the circuit lattice, 
matrix $L=diag(L_{A},L_{B})$, $\mathcal V=(\mathcal V_{A},\mathcal V_{B})^{T}$ is the Bloch states to describe electric potential distributions, and the Pauli
matrices are for the space spanned by $(\mathcal V_{A},\mathcal V_{B})$. 
The coefficients in front of the Pauli matrices are given as\\
\begin{equation}
d_{0}(\bm{k})= C_{1}+C_{2}+C_{3}+C_{4}+(C_{GA}+C_{GB})/2+(C_{A}+C_{B})(1-cosk_{c}),\label{eq:d0}
\end{equation}
\begin{equation}
d_{1}(\bm{k})= -C_{1}-C_{2}cos(k_{b}-k_{a})-C_{3}cosk_{a} -C_{4}cos(k_{c}-k_{a}),\label{eq:dxx}
\end{equation}
\begin{equation}
d_{2}(\bm{k})= C_{2}sin(k_{b}-k_{a})-C_{3}sink_{a}+C_{4}sin(k_{c}-k_{a}),\label{eq:dy}
\end{equation}
\begin{equation}
d_{3}(\bm{k})=(C_{GA}-C_{GB})/2+(C_{A}-C_{B})(1-cosk_{c}).\label{eq:dz}
\end{equation}
To simplify the calculations, we set $L_{A}=L_{B}=L_{G}$, therefore
the matrix $L$ in equation (\ref{eq:Hpsi}) is proportional to an identity
matrix.
Without this simplification, we will have a generalized eigenvalue problem at hand, showing similar bands and the same physics.
Solving equation (\ref{eq:Hpsi}), we obtain two branches of dispersions 
$\omega_{1,2}^{-2}(\bm{k})=L_{G}\left[d_{0}(\bm{k})\pm\sqrt{d_{1}^{2}(\bm{k})+d_{2}^{2}(\bm{k})+d_{3}^{2}(\bm{k})}\right]$ in the BZ.
By varying the parameters of capacitor $C_i$s, we can artificially tune the dispersion
of $\omega_{1,2}^{-2}(\bm{k})$ and arrive at the desired bands structure. 
In the following section, we will
show that the nodal-ring-type and Weyl point-type bands touching 
points 
are available in our circuit lattice.

\section{Nodal-line and drumhead-like surface state}
\begin{figure}[h]
\includegraphics[width=0.9\textwidth]{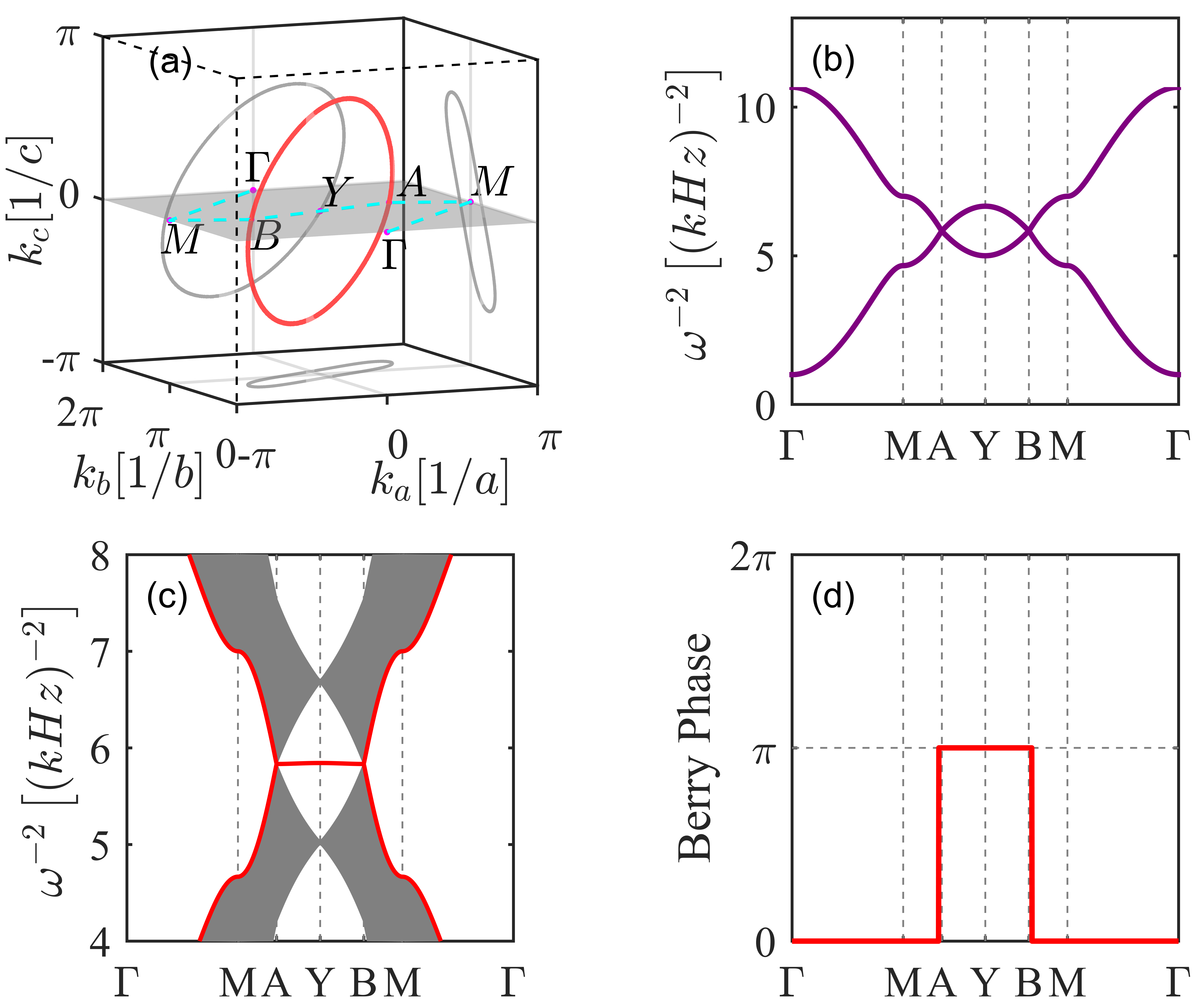}
\caption{(a) Nodal-line (in red) in the BZ and its projection on
the (001), (010) and (100) planes (in grey). The parameters are set as 
$C_{1}=1mF$, 
$C_{2}=2mF$, 
$C_{3}=1mF$, 
$C_{4}=0.833mF$, 
$C_{GA}=C_{GB}=1mF$, and 
$L_{A}=L_{B}=1mH$. 
(b) Band structure along $\Gamma-X-A-B-X-\Gamma$, where A and B are
two points with $k_{c}=0$ on the nodal-line as labeled in (a). 
(c) The band dispersions with the surface states (red colour lines) on the
(001) surface. The drumhead-like surface states are nestled inside
the projection of the nodal ring. 
(d) The Berry phase $\theta_{k_{\parallel}}$
equals $\pi$ for $k_{\parallel}$ inside the nodal ring, while it
is zero for $k_{\parallel}$ outside the nodal ring. }
\end{figure}

The closed loops of bands crossing points can be protected
by the coexistence of space-inversion symmetry and time-reversal symmetry\cite{PhotonicTI_LingLu_NPgyroid_2013}. The LC circuit is naturally time-reversal symmetric,
therefore we only need to choose a group of parameters of the inductors
and capacitors to preserve the space-inversion symmetry. For the circuit
network given in Fig.~1(b), the space-inversion symmetry
can be obtained by setting $C_{GA}=C_{GB}$, with
the inversion center located at the middle point of node A and node B.
In this case, $d_{3}(\bm{k})$ vanishes ($C_A=C_B=0$ in Fig.~1(b)). 
The conditions for bands degeneracy require $d_{1}(\bm{k})=0$ and $d_{2}(\bm{k})=0$
to be satisfied simultaneously, leading to two restrictions for three variables $\bm{k}=(k_{a},k_{b},k_{c})$, which gives
one-dimensional solution space and form a continuous nodal-ring in the $\bm{k}$ space.
Fig.~2(a) illustrates 
numerically obtained nodal-ring structure centered at $Y=(0,\pi,0)$ point. The band dispersions along 
$\Gamma$-$M$-$A$-$Y$-$B$-$M$-$\Gamma$ are
shown in Fig.~2(b), where A and B are two nodal points
in $k_{c}=0$ plane.

The topological properties of a nodal-line can be inferred from the winding number $(-1)^{\zeta_{1}}=\frac{1}{\pi}\oint_{C}\mathcal{A}(\bm{k})\cdot d\bm{k}$,
where $\mathcal{A}(\bm{k})$ is the Berry connection and $C$ is a closed
loop in the momentum space pierced by the nodal-line.
$\zeta_{1}=1$ in our model means that the nodal-line is topologically stable. 
Due to bulk-boundary correspondence, the non-trivial nodal-line structure indicates a novel surface state. 
Based on the tight-binding-like model~\ref{eq:H_bulk}, the surface state in the (001) direction is calculated
as shown in Fig.~2(c), where a flat surface band nestles
inside of the projected node-ring.
In order to relate the non-trivial topology induced by the bulk band
singularity to edge modes, 
we calculate the Berry phase of the one-dimensional
systems $H_{k_{\parallel}}(k_c)$ parameterized by the in-plane momentum $k_{\parallel}=(k_{a},k_{b})$.
The Berry phase for the one-dimensional system is defined as $\theta_{k_{\parallel}}=\int\mathcal{A}_{k_{\parallel}}dk_{\perp}$, where
$\mathcal{A}_{k_{\parallel}}$ is the Berry connection matrix  defined as
$\mathcal{A}_{k_{\parallel}}=\langle \mathcal{V}_{1}(\bm{k})|i\partial_{k_{c}}|\mathcal{V}_{1}(\bm{k})\rangle$.
In the one-dimensional parametrized systems, the Berry phase equals $\pi$ for $k_{\parallel}$ inside the nodal-ring,
while it is zero for $k_{\parallel}$ outside the nodal-ring as shown
in Fig.~2(d).

\section{Weyl points and surface `Fermi-arc'}
\begin{figure}[ht]
\includegraphics[width=0.9\textwidth]{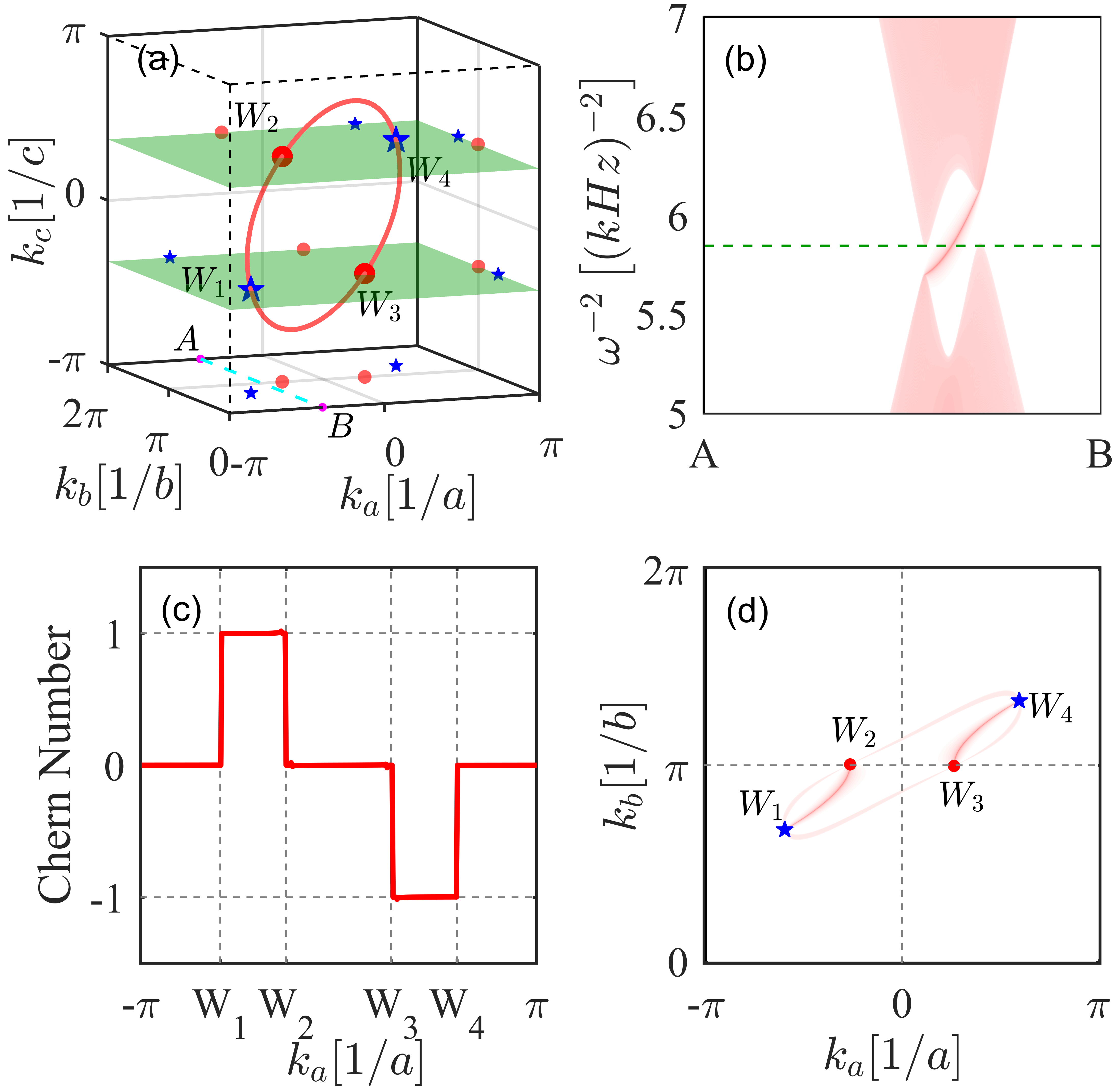}
\caption{(a) Four Weyl points in the Brillouin zone and their
projections on (001), (010) and (100) direction. 
$C_{A}=0.2mF$, $C_{B}=0.01mF$ and $C_{GA}=0.77mF$ are used in the calculations.
The other parameters are the same as figure 2. 
The positions of the Weyl points are the intersection points
between the nodal-ring determined by $d_{1}(\bm{k})=d_{2}(\bm{k})=0$ and the two
planes determined by $d_{3}(\bm{k})=0$. The chirality are indicated as blue
stars for $\chi=+1$ and red points for $\chi=-1$. 
(b) The gapless surface band dispersions on the A-B-$k_c$ plane and terminates in the (001) direction. 
(c) The Chern numbers for the two-dimensional planes perpendicular to $k_a$.
Moving along $k_{a}$, the  Chern number increase (decrease) when the plane passing through the Weyl points with $+1$ ($-1$) chirality.
(d) On the (001) surface, Fermi arcs connect the
projections of the bulk Weyl nodes of opposite chiralities onto the
surface.}
\end{figure}

Starting with nodal-line states, Dirac states or Weyl states 
are possible to emerge by introducing
symmetry breaking terms\cite{PhotonicTI_LingLu_NPgyroid_2013,Weyl_HMWeng_PRX_2015,YR_PRL_2015,YR_PRL_2017}.
Here we show the Weyl states can be realized in the circuit lattice shown in Fig.~1(c, d)
. The space inversion
symmetry of the circuit lattice is easy to be removed by setting $C_{GA}\ne C_{GB}$ and $C_{A}\ne C_{B}$.  In this case, $d_{1}(\bm{k})$ and $d_{2}(\bm{k})$ are not affected and the nodal-ring structure as solutions of $d_{1}(\bm{k})=d_{2}(\bm{k})=0$ remains. 
Now the gap closing condition further requires $d_{3}(\bm{k})=0$, 
which leads to 
${\rm cos}k_c=1+(C_{GA}-C_{GB})/2(C_A-C_B)$. 
Tuning the parameters of $C_{GA}$, $C_{GB}$, $C_A$, and $C_B$ to make the absolute value of the right-hand side of
the equation less than 1, $d_{3}(\bm{k})=0$ determines two planes perpendicular to $k_{c}$ axis in the first BZ.
When these two planes intersect with the nodal-ring, as shown
in Fig.~3(a), there are four crossing
points that are the desired Weyl points in the Brillouin zone. 
Because time-reversal symmetry maps Weyl
point at $\bm{k}$ to $-\bm{k}$ with the same chirality,
to neutralize the chirality in the BZ there must be two other Weyl points with opposite chirality. Therefore the minimal number of Weyl
points in our circuit lattice has to be four. 
The chiralities of the Weyl points are determined in the following way. 
We calculate the Chern number for the lower bulk bands on the planes perpendicular to 
$k_{a}$ axis in the 3D BZ. 
As shown in Fig.~3(c), moving along $k_{a}$, the increasing (decreasing) of Chern number
when the plane passing through the Weyl points indicates that the chiral value of the Weyl points are $+1$ ($-1$).
The Weyl points with $+1$ chirality are marked with blue star while red point for $-1$ chirality in Fig.~3(a).   

As a result of the non-zero Chern numbers,
topologically protected gapless chiral surface states emerge in
the band gap away from the Weyl points. An example of the non-trivial
gapless surface band dispersions, on the $A$-$B$-$k_c$ plane between 
$W_1$ and $W_2$ points and terminates in the (001) direction are shown in Fig.~3(b). 
A surface mode profile at the frequency of the Weyl points is also shown
in Fig.~3(d).
Similar to the Fermi arcs in Weyl semimetals, two line-segment-like 
surface dispersions connect Weyl pairs with opposite
chirality are clearly shown in the surface Brillouin zone.
The surface `Fermi arc' for (100) and (010) directions surface can be calculated
with the similar method.

\section{Stability of the Weyl points}

\begin{figure}[ht]
\includegraphics[width=0.9\textwidth]{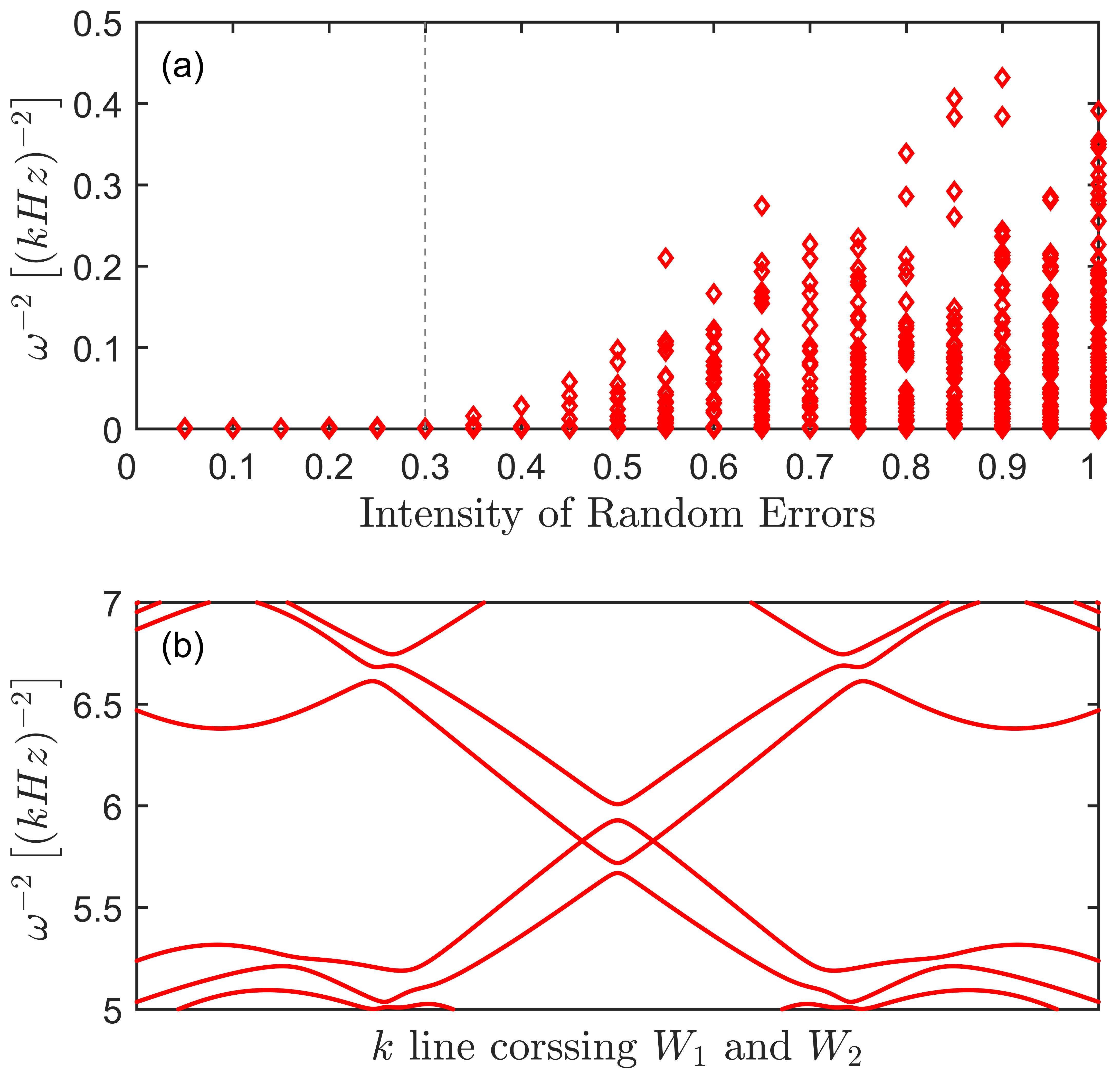}
\caption{(a) The band gap, $min[\omega_{N/2+1}^{-2}-\omega_{N/2}^{-2}]$ as a function of the tolerance values for a $3\times3\times3$ super cell. $N$ is number of total bands for the $3\times3\times3$ super cell. The parameters are the same as figure 3 without considering the fabrication errors. After taking the fabrication errors into consideration, we repeat 100 times calculations for each tolerance values.
The numerical results show that the gap equals zero, i.e., Weyl points exist for the tolerance values less than 30\%. Exceeds this value, the gap opens by the fabrication errors.  
(b) The band dispersions along the $\bm k$ line that 
crossing two Weyl points for a $3\times3\times3$ 
super cell with $\pm 20\%$ tolerance values on the capacitors.}
\end{figure}

In the above discussions, the inductance and capacitance in 
the circuit lattice are proposed as
a set of precise values, but the fabrication error of the electronic
devices bring about certain range of tolerance values.
In this paragraph, we discuss whether the fabrication errors influence the results given above. For the nodal-line state,
maintaining intrinsic space inversion symmetry is a necessary condition.
When the fabrication errors are taken into consideration, intrinsic space-inversion symmetry is too demanding to preserve, therefore nodal-line becomes unstable.
However, the existence of the Weyl points do
not require extra symmetries except the discrete translation invariant symmetry. It is expected to be
more stable than the nodal-line state under the perturbation of the fabrication errors.
In order to investigate the stability of 
the Weyl points in the circuit lattice,
we employ a $3\times3\times3$ super cell and add random numbers within certain range to the parameter of capacitors. 
In the language of solid state physics, the random errors 
perturb hopping terms and on-site
energies in eqs. (6-9) shown in the supplementary material.
To simplify the calculation without loss of key points, 
we keep $L_{A}$ and $L_{B}$ fixed.
For each tolerance value, we repeat calculations 100 times 
with different random errors and find the minimal value of the 
gap between band $N/2+1$ and $N/2$ in each time. Here, $N$ is the number of total bands for the 
$3\times3\times3$ super cell. 
As long as these two bands are not gaped, i.e., the gap is zero, we could safely say that the Weyl points still survive.
The numerical results in Fig.~4(a) show that the Weyl points exist for tolerance values less than 30\%. Exceeds this value, the gap opens and we get a topologically trivial circuit. 
Take a super cell with $\pm20\%$  tolerance values for example, we calculate 
its band structures along the $\bm k$ line crossing two of the gap closing points.
The band dispersions in Fig.~4(b) show that two bands cross each other 
and form a pair of Weyl points near $\omega^{-2}=5.8 {(kHz)}^{-2}$.

\section{Conclusions} 
To summarize, we report that the topological nodal-line state and 
Weyl state can be realized in a three-dimensional classical circuit system.
We derived a two-by-two tight-binding-like model to investigate its topological nature. 
Based on this model, we show that the nodal-line structure is protected by the 
inversion symmetry, which can be achieved by setting $C_{GA}=C_{GB}$ and $C_{A}=C_{B}$. Break the inversion symmetry, gap opens along the nodal-line, while leaves
four Weyl points in the first BZ. We check that the Weyl state is robust with the fabrication error of the electrical devices.
The key characters of the above two topological state,
namely the drumhead-like surface states for nodal-line state and
`Fermi-arcs' surface state for Weyl state are conformed by the tight-binding-like model terminated on the (001) direction surface.
Moreover, the topological phase transition between nodal-line state and Weyl state can be tuned by changing the parameters of the capacitors and inductors
or by controlling the connecting or disconnecting of some elements, for example the capacitors $C_{A, B}$ and $C_{GA, GB}$ in Fig. 1(c), in the circuit lattice.
In the experimental aspect, the proposed circuit lattice 
can be manufactured with current technology. The topological properties, such as
the patterns for the `Fermi arcs' connecting the Weyl point pairs,
can be detected by measuring the potential distribution on the surface of the circuit lattice.
This work offers a new, robust platform for realizing and tuning 
topological nodal-line state and Weyl state in the classical system.

\noindent \textbf{Acknowledgements} The authors thank Hua Jiang, Yuanyuan Zhao and Ang Cao for very helpful discussions. This work was supported by the {National Key Research and Development Program of China} (No.~2017YFA0303402, No.~2017YFA0304700, and No.~2016YFA0300600), the National Natural Science Foundation
of China (No.~11674077, No.~11422428,  No.~11674369, and No.11404024). 
R. Y. acknowledges funding from the National Thousand Young Talents Program. 
The numerical calculations in this work have been done on the supercomputing system in the Supercomputing Center of Wuhan University.

\section{Supplementary Material}

In this section, we present the details of the methods used to calculate the band structure of the circuit lattice.
We start by deriving the tight-binding-like model given in equation~(1). The
current $I$ passing through a two-terminal circuit element for a given
drop in voltage $\Delta v$ is described by the Ohm's law $I=Y\Delta v$, where
$Y$ is admittance. In an alternating current (AC) sinusoidal circuits,
the admittance for the ideal inductors and capacitors are $Y_{L}=1/j\omega L$
and $Y_{C}=j\omega C$, respectively. $\omega$ is the frequency for
the sinusoidal signal and we use the imaginary unit $j=\sqrt{-1}$, following
the convention in electric circuits. The currents flow into nodes
A and B in the cell located at $\boldsymbol{R}=\bm{0}$ of circuit
lattices given in Fig.~1(c, d) of main text are given as
\begin{align}
I_{A}(\bm 0)= & j\omega\big(C_{1}[v_{B}(\bm 0)-v_{A}(\bm 0)]+C_{2}[v_{B}(\bm b-\bm a)-v_{A}(\bm 0)]+C_{3}[v_{B}(-\bm a)-v_{A}(\bm 0)]\nonumber \\
 & +C_{4}[v_{B}(\bm c-\bm a)-v_{A}(\bm 0)]+C_{A}[v_{A}(\bm c)-v_{A}(\bm 0)]+C_{A}[v_{A}(-\bm c)-v_{A}(\bm 0)]\nonumber \\
 & +C_{GA}[0-v_{A}(\bm 0)]\big)+[0-v_{A}(\bm 0)]/(j\omega L_{A})\nonumber \\
= & j\omega\big(C_{1}v_{B}(\bm 0)+C_{2}v_{B}(\bm b-\bm a)+C_{3}v_{B}(-\bm a)+C_{4}v_{B}(\bm c-\bm a)+C_{A}v_{A}(\bm c)\nonumber \\
 & +C_{A}v_{A}(-\bm c)-(C_{GA}+C_{1}+C_{2}+C_{3}+C_{4}+2C_{A})v_{A}(\bm 0)\big)-v_{A}(\bm 0)/(j\omega L_{A}),
 \label{eq:IA}
\end{align}
and
\begin{align}
I_{B}(\bm 0)= & j\omega\big(C_{1}[v_{A}(\bm 0)-v_{B}(\bm 0)]+C_{2}[v_{A}(\bm a-\bm b)-v_{B}(\bm 0)]+C_{3}[v_{A}(\bm a)-v_{B}(\bm 0)]\nonumber \\
 & +C_{4}[v_{A}(\bm a-\bm c)-v_{B}(\bm 0)]+C_{B}[v_{B}(\bm c)-v_{B}(\bm 0)]+C_{B}[v_{B}(-\bm c)-v_{B}(\bm 0)]\nonumber \\
 & +C_{GB}[0-v_{B}(\bm 0)]\big)+[0-v_{B}(\bm 0)]/(j\omega L_{B})\nonumber \\
= & j\omega\big(C_{1}v_{A}(\bm 0)+C_{2}v_{A}(\bm a-\bm b)+C_{3}v_{A}(\bm a)+C_{4}v_{A}(\bm a-\bm c)+C_{B}v_{B}(-\bm c)\nonumber \\
 & +C_{B}v_{B}(\bm c)-(C_{GB}+C_{1}+C_{2}+C_{3}+C_{4}+2C_{B})v_{B}(\bm 0)\big)-v_{B}(\bm 0)/(j\omega L_{B}).\label{eq:IB}
\end{align}
The Kirchhoff\textquoteright s current law demands that $I_{A}(\bm 0)$
and $I_{B}(\bm 0)$ are zero. Divide $j\omega$ on both sides of equation (\ref{eq:IA})
and (\ref{eq:IB}), we get
\begin{align}
-\bigg(C_{1}v_{B}(\bm 0)+C_{2}v_{B}(\bm b-\bm a)+C_{3}v_{B}(-\bm a)+C_{4}v_{B}(\bm c-\bm a)+C_{A}v_{A}(\bm c)\nonumber \\
+C_{A}v_{A}(-\bm c)-(C_{GA}+C_{1}+C_{2}+C_{3}+C_{4}+2C_{A})v_{A}(\bm 0)\bigg) & =\frac{1}{\omega^{2}L_{A}}v_{A}(\bm 0),\label{eq:HA}
\end{align}
\begin{align}
-\bigg(C_{1}v_{A}(\bm 0)+C_{2}v_{A}(\bm a-\bm b)+C_{3}v_{A}(-\bm b)+C_{4}v_{A}(\bm a-\bm c)+C_{B}v_{B}(-\bm c)\nonumber \\
+C_{B}v_{B}(\bm c)-(C_{GB}+C_{1}+C_{2}+C_{3}+C_{4}+2C_{B})v_{B}(\bm 0)\bigg) & =\frac{1}{\omega^{2}L_{B}}v_{B}(\bm 0).\label{eq:HB}
\end{align}
Using the similar method, we can obtain the equations for the potential
distribution $v_{A}(\boldsymbol{R})$ and $v_{B}(\boldsymbol{R})$
on the whole lattice. Written these equitations to a matrix form, we have
\begin{equation}
YV=\frac{1}{\omega^{2}L}V,\label{eq:Htb}
\end{equation}
where $V$ is a vector for the potential distribution on A and B nodes
of the circuit lattice, $L$ is a diagonal matrix composed of $L_{A}$
and $L_{B}$ and $Y$ is the admittance matrix containing all information of the involving capacitors in our circuit. 
The eigenvalue-equation-like form of equation~\ref{eq:Htb} reminds us of the stationary Schroedinger equation in quantum mechanics, in which $V$ is the counterpart of wave function and $1/\omega^2L$ plays the role of energy eigenvalue. 
Following this insight, the admittance matrix $Y$ can be interpreted as tight-binding Hamiltonian in real space.
Hopping terms and on-site energies of the tight-binding
model can be extracted from equation~\ref{eq:HA} and \ref{eq:HB}, which are listed below. %
\begin{equation}
H_{AB}(\bm R=\bm 0)=-C_{1};\,\;H_{AB}(\bm R=\bm b-\bm a)=-C_{2};\;\;H_{AB}(\bm R=-\bm a)=-C_{3};\label{eq:xx1}
\end{equation}
\begin{equation}
H_{AB}(\bm R=\bm c-\bm a)=-C_{4};\,\;H_{AA}(\bm R=\pm \bm c)=-C_{A};\;\;H_{BB}(\bm R=\pm \bm c)=-C_{B};\label{eq:xx2}
\end{equation}
\begin{equation}
H_{AA}(\bm R=\bm 0)=C_{GA}+C_{1}+C_{2}+C_{3}+C_{4}+2C_{A};\label{eq:HAAR}
\end{equation}
\begin{equation}
H_{BB}(\bm R=\bm 0)=C_{GB}+C_{1}+C_{2}+C_{3}+C_{4}+2C_{B},\label{eq:HBBR}
\end{equation}
where $\bm R$ is the lattice vector, $H_{mn}(\bm R)$ are the tight-binding parameters
between node $m$ located at the home unit cell and node $n$ located at $\bm R$.
With these terms, the tight-bind Hamiltonian in the momentum $\boldsymbol{k}$
space can be obtained through Fourier transformation $H_{mn}(\bm k)=\sum_{\boldsymbol{R}}e^{i\boldsymbol{k}\cdot\boldsymbol{R}}H_{mn}(\boldsymbol{R})$.
The elements of the Hamiltonian are given as %
\begin{equation}
H_{AA}(\bm {k})=C_{GA}+C_{1}+C_{2}+C_{3}+C_{4}+2C_{A}-C_{A}(e^{ik_{c}}+e^{-ik_{c}}),\label{eq:HAAk}
\end{equation}
\begin{equation}
H_{BB}(\boldsymbol{k})=C_{GB}+C_{1}+C_{2}+C_{3}+C_{4}+2C_{B}-C_{B}(e^{ik_{c}}+e^{-ik_{c}}),\label{eq:HBBk}
\end{equation}
\begin{equation}
H_{AB}(\boldsymbol{k})=-C_{1}-C_{2}e^{i(k_{b}-k_{a})}-C_{3}e^{-ik_{a}}-C_{4}e^{i(k_{c}-k_{a})},\label{eq:HABk}
\end{equation}
with the Bloch like basis function $|\text{\ensuremath{\mathcal V}}_{m}(\bm k)\rangle=\sum_{\bm R}e^{i\boldsymbol{k}\cdot\boldsymbol{R}}|v_{m}(\bm R)\rangle,$
$m=A,B$. Rewrite the 2 by 2 Hamiltonian matrix with the help of Pauli
matrices, we get %
\begin{equation}
H(\boldsymbol{k})=\sum_{i=0}^{3}d_{i}(\bm k)\sigma_{i}
\end{equation}
with
\begin{equation}
d_{0}(\bm k)=C_{1}+C_{2}+C_{3}+C_{4}+(C_{GA}+C_{GB})/2+(C_{A}+C_{B})(1-cosk_{c}),\label{eq:d0}
\end{equation}
\begin{equation}
d_{1}(\bm k)=-C_{1}-C_{2}cos(k_{b}-k_{a})-C_{3}cosk_{a}-C_{4}cos(k_{c}-k_{a}),\label{eq:d1}
\end{equation}
\begin{equation}
d_{2}(\bm k)=C_{2}sin(k_{b}-k_{a})-C_{3}sink_{a}+C_{4}sin(k_{c}-k_{a}),\label{eq:d2}
\end{equation}
\begin{equation}
d_{3}(\bm k)=(C_{GA}-C_{GB})/2+(C_{A}-C_{B})(1-cosk_{c}),\label{eq:d3}
\end{equation}
which is the tight-binding-like model given in equation~2.

For the single layer honeycomb circuit lattice given in Fig.~1(a), $C_4$, $C_A$, and $C_B$ vanish. If we set $C_{GA}=C_{GB}$, we have  
\begin{equation}
d_{1}(\bm k)=-C_{1}-C_{2}cos(k_{b}-k_{a})-C_{3}cosk_{a},\label{eq:d10}
\end{equation}
\begin{equation}
d_{2}(\bm k)=C_{2}sin(k_{b}-k_{a})-C_{3}sink_{a},\label{eq:d20}
\end{equation}
and
\begin{equation}
d_{3}(\bm k)=0.\label{eq:d30}
\end{equation}
The appearance of gap closing points requires $d_{1}(\bm k)=d_{2}(\bm k)=0$, which lead to
\begin{equation}
C_1+C_2e^{-i(k_a-k_b)}+C_3e^{-ik_a}=0.\label{eq:d30}
\end{equation}
As long as the three values of $C_1$, $C_2$ and $C_3$ satisfy the triangle inequality theorem, two band-crossing points can be found in the $k_a$-$k_b$ plane.

\bibliography{refs}
\bibliographystyle{apsrev4-1}

\end{document}